\documentclass[acmsmall,screen]{acmart}

\usepackage{amsmath,amsfonts}
\usepackage{algorithmic}
\usepackage{algorithm}
\usepackage{array}
\usepackage[caption=false,font=normalsize,labelfont=sf,textfont=sf]{subfig}
\usepackage{textcomp}
\usepackage{stfloats}
\usepackage{url}
\usepackage{verbatim}
\usepackage{graphicx}

\usepackage{listings}
\usepackage{subfig}
\usepackage{graphicx}
\usepackage{multirow}
\usepackage{tabularx}
\usepackage{xcolor}
\usepackage{rotating}
\usepackage{booktabs}
\usepackage{balance}

\usepackage{float} 
\usepackage{graphicx} 
\usepackage{caption} 
\usepackage{subcaption} 
\usepackage[symbol]{footmisc}
\usepackage{threeparttable}  
\usepackage{array}  
\usepackage{url}  
\usepackage{xcolor}
\usepackage{colortbl}
\usepackage{scalerel,graphicx,xparse}

\usepackage{enumitem}
\usepackage{orcidlink}

\newtheorem{hyp}{Hypothesis}
\newtheorem{reco}{Recommendation}

\AtBeginDocument{%
  }

\setcopyright{acmlicensed}
\copyrightyear{2018}
\acmYear{2018}
\acmDOI{XXXXXXX.XXXXXXX}

\acmJournal{JACM}
\acmVolume{37}
\acmNumber{4}
\acmArticle{111}
\acmMonth{8}

\begin{document}

\title{Human, AI, and Hybrid Ensembles for Detection of Adaptive, RL-based Social Bots}


\author{Valerio La Gatta}
\email{valerio.lagatta@northwestern.edu}
\orcid{0000-0002-5941-4684}
\affiliation{%
  \institution{Northwestern University}
  \city{Evanston}
  \state{Illinois}
  \country{USA}
}

\author{Nathan Subrahmanian}
\email{nsubrahmanian@brandeis.edu}
\orcid{0000-0003-3112-8253}
\affiliation{%
  \institution{Brandeis University}
  \city{Boston}
  \state{Massachusetts}
  \country{USA}
}

\author{Kaitlyn Wang}
\email{kaitlynwang2027@u.northwestern.edu}
\affiliation{%
  \institution{Northwestern University}
  \city{Evanston}
  \state{Illinois}
  \country{USA}
}

\author{Larry Birnbaum}
\email{l-birnbaum@northwestern.edu}
\orcid{0000-0002-9315-2796}
\affiliation{%
  \institution{Northwestern University}
  \city{Evanston}
  \state{Illinois}
  \country{USA}
}

\author{V.S. Subrahmanian}
\email{vss@northwestern.edu}
\orcid{0000-0001-7191-0296}
\affiliation{%
  \institution{Northwestern University}
  \city{Evanston}
  \state{Illinois}
  \country{USA}
}

\renewcommand{\shortauthors}{La Gatta et al.}

\begin{abstract}
    The use of reinforcement learning to dynamically adapt and evade detection is now well-documented in several cybersecurity settings including Covert Social Influence Operations (CSIOs), in which bots try to spread disinformation. While AI bot detectors have improved greatly, they are largely limited to detecting static bots that do not adapt dynamically. We present the first systematic study comparing the ability of humans, AI models, and hybrid Human-AI ensembles 
    in detecting \emph{adaptive} bots \emph{powered by reinforcement learning}. Using data from a controlled, IRB-approved, five-day experiment with participants interacting on a social media platform infiltrated by RL-trained bots spreading disinformation to influence participants on 4 topics, we examine factors potentially shaping human detection capabilities: demographic characteristics, temporal learning effects, social network position, engagement patterns, and collective intelligence mechanisms. 
    We first test 13 hypotheses comparing human bot detection performance against state-of-the-art AI approaches utilizing both traditional machine learning and large language models. We further investigate several aggregation strategies that combine human reports of bots with AI predictions, as well as retraining protocols that leverage human supervision. Our findings challenge intuitive assumptions about bot detection, reveal unexpected patterns in how humans identify bots, and show that combining human bot reports with AI predictions outperforms humans alone and AI alone. We conclude with a discussion of the practical implications of these results for industry. 
\end{abstract}



\received{22 March 2026}

\maketitle

\section{Introduction}

Covert Social Influence Operations (CSIOs) threaten democratic discourse by leveraging social media to manipulate information ecosystems, spread disinformation, and polarize communities~\cite{mannocci2024detection,martin2019recent,zannettou2019let}. 

However, a common theme in cybersecurity is that attackers adapt. This occurs with CSIOs as well: state-backed cyber threat actors deploy increasingly sophisticated strategies that adapt dynamically to countermeasures~\cite{cima2024coordinated,alizadeh2020content}.  Recent advances in artificial intelligence (AI) have accelerated this threat. Reinforcement learning (RL) agents can dynamically adapt to static malware detectors
\cite{anderson2018learning,song2020mab}, create attack paths through a network
\cite{lee2025approach}, neutralizing graph-neural network based detectors focused on node injection attacks\cite{sun2020adversarial}, and creating fake reviews in online marketplaces\cite{zhang2023sockdef}. 

\cite{10.1145/3696410.3714729} presents a reinforcement learning-based CSIO (RL\_CSIO) attack in which an adversary can deploy autonomous, dynamically adaptive agents that control bot accounts to maximize influence while changing their behavior to evade detection~\cite{10.1145/3696410.3714729}. RL-operated bots, unlike static accounts, learn adaptive behaviors through continuous interaction, developing sophisticated strategies that simultaneously spread narratives and evade detection by AI classifiers. This potentially marks a fundamental shift from manually orchestrated campaigns run by nation states, to a future in which nation states and other malign actors may well leverage reinforcement learning to create autonomous, self-optimizing systems capable of real-time adaptation to platform defenses. Such RL\_CSIO's will continuously learn and exploit a model of the classifiers used by social platforms --- and perhaps others --- to detect bots. Today, most bot detectors in industry use a combination of rule-based and machine learning classifiers\footnote{\url{https://www.snopes.com/articles/435482/spot-a-bot/}}\footnote{\url{https://mitsloan.mit.edu/ideas-made-to-matter/study-finds-bot-detection-software-isnt-accurate-it-seems}} that attempt to learn a model that distinguishes bots from genuine accounts. However if the adversary uses RL to dynamically change its behavior, this becomes challenging as this attack model changes continuously.

There has been substantial research in automated bot detection. AI approaches have evolved from early classifiers based on behavioral features~\cite{yang2019arming,cresci2020decade} to sophisticated systems leveraging graph neural networks~\cite{huang2025semi,10549990}, large language models~\cite{zhou2025lgb,10924316}, and temporal models~\cite{yang2023social}. Detection frameworks specifically targeting CSIO-operated accounts employ content analysis~\cite{alizadeh2020content,addawood2019linguistic}, behavioral pattern recognition~\cite{DBLP:journals/corr/abs-1804-05232,cresci2016dna,luceri2020detecting}, and coordinated signal detection~\cite{sharma2021identifying,luceri2024unmasking,pacheco2020uncovering}. Despite this progress, automated methods face inherent limitations: they operate retrospectively, they require labeled training data that may not generalize to new evasion tactics, and they are engaged in an asymmetric arms race where adaptive bots can learn to evade known detectors.

In contrast, human detection remains critically understudied, yet may offer complementary capabilities. The limited research on human-based bot detection reveals problematic patterns: individuals frequently fail to recognize sophisticated bots, achieving precision typically below 0.5~\cite{kolomeets2024experimental}. Paradoxically, greater social media experience does not guarantee improved accuracy and may foster overconfidence~\cite{kenny2024duped}. Cognitive and political biases further reduce sensitivity to bots expressing aligned viewpoints~\cite{doi:10.1177/1461444820942744,10.1145/3411764.3445109}. However, these findings emerge almost exclusively from studies of \emph{static} bot behaviors where the bots do not adapt dynamically. Critical questions remain unanswered: \emph{How do humans perform against adaptive, RL-based bots that actively learn to evade detection? What individual, social, and collective factors shape detection capabilities in adversarial contexts? Can human judgment be aggregated to overcome individuals' limitations? {Can human reports and AI predictions be combined to improve detection performance?}}

This paper addresses these questions through the first controlled investigation of human bot detection against RL-based bots\footnote{Hereafter, we use the term "bots" to refer to RL-based adaptive bots proposed by \cite{10.1145/3696410.3714729}.} operating in active CSIOs. We examine whether demographic characteristics are linked to users' detection performance, whether users learn to identify bots over time, how social activity influences accuracy, and whether collective intelligence mechanisms can aggregate individual judgments effectively.
{ 
We benchmark human capabilities against three state-of-the-art AI detectors to assess complementarity, and investigate hybrid strategies that combine human reports with AI predictions. }

Our analysis leverages data from a five-day controlled experiment in which 225 participants interacted on a social media platform infiltrated by four concurrent RL\_CSIO campaigns~\cite{10.1145/3696410.3714729}, providing unprecedented insight into human detection capabilities under realistic adversarial conditions. Our contributions are:

\begin{itemize}[leftmargin=*]
\item  We provide the first systematic analysis of human-based bot detection performance against \emph{RL-based adaptive bots} in active influence operations, building on prior work on human bot detection ability against \emph{static, non-adaptive bots} \cite{cresci2017paradigm,kolomeets2024experimental,martini2021bot,kats2022have}

\item  We test 13 hypotheses examining how individual characteristics (age, education, language proficiency, platform usage), temporal dynamics (learning effects over time), and collective mechanisms (multiple report aggregation) shape detection performance under realistic adversarial conditions.

\item We compare human ensemble performance against three recent AI detectors spanning traditional machine learning and large language models, examining both overall performance metrics and pairwise agreement patterns to assess whether humans and AI identify bots through similar or distinct mechanisms.

\item {  We investigate aggregation strategies that combine human reports with AI predictions and retraining protocols that leverage human supervision to improve detector performance.}
\end{itemize}

\section{Related Work}

\noindent\textbf{Influence Campaigns on Social Media.}
Covert social influence operations (CSIOs) represent a growing threat to democratic discourse, increasingly leveraging sophisticated, multi-platform coordination strategies~\cite{cima2024coordinated,mannocci2024detection}. Between 2013 and 2018, at least 53 covert campaigns were documented across 24 countries~\cite{martin2019recent}, with state-backed actors continually adapting to platform countermeasures~\cite{zannettou2019let}. The emergence of reinforcement learning (RL) has further advanced these operations, enabling bots to strategically balance influence maximization and detection avoidance~\cite{10.1145/3696410.3714729}.  
A wide range of AI detection approaches have been proposed, targeting both automated accounts~\cite{DBLP:journals/corr/abs-1804-05232,cresci2016dna} and human operators~\cite{ferrara2023social}. These methods include content-based~\cite{alizadeh2020content,addawood2019linguistic}, behavior-based~\cite{luceri2020detecting,sharma2021identifying}, and hybrid frameworks~\cite{subrahmanian2016darpa,10.1145/3394231.3397889}. Recent advances integrate multiple behavioral signals such as timing synchrony and coordination~\cite{luceri2024unmasking}, while unsupervised methods improve adaptability to unseen scenarios~\cite{pacheco2020uncovering}. Yet, these AI methods remain constrained by retrospective analyses and limited generalization to evolving CSIO tactics.  

In contrast, bot detection by humans has received far less attention. Prior work primarily explores public awareness~\cite{liberg2021risk} and training interventions such as ``spot the troll'' exercises~\cite{10.1093/pnasnexus/pgad094}, focusing on static bot behaviors. \emph{This study bridges this gap by presenting the first systematic investigation of real-time human detection of adaptive, RL-driven bots embedded in active influence campaigns, examining how demographic, temporal, and collective factors shape human detection performance.}

\noindent \textbf{Bot Detection.}
Social bot detection research has progressed from early classifiers based on linguistic, behavioral, and network features~\cite{yang2019arming,cresci2020decade,yang2023social} to advanced models leveraging graph neural networks~\cite{huang2025semi,10549990} and large language models~\cite{zhou2025lgb,10924316}. Tools such as Botometer~\cite{DBLP:journals/corr/abs-2201-01608} remain widely used, yet comparative studies show low agreement among detectors, with inconsistent classifications across identical datasets~\cite{martini2021bot}.  

Human performance in bot detection, by contrast, remains poorly understood. Empirical studies reveal that individuals often fail to recognize sophisticated bots~\cite{cresci2017paradigm}, achieving precision typically below 0.5 and highly variable recall~\cite{kolomeets2024experimental,martini2021bot,kats2022have}. Moreover, greater social media experience does not guarantee improved accuracy and may instead foster overconfidence~\cite{kenny2024duped}. Cognitive and political biases also reduce sensitivity to bots expressing opposing viewpoints~\cite{doi:10.1177/1461444820942744,10.1145/3411764.3445109}. Some evidence, however, suggests gradual learning effects with repeated exposure~\cite{10.1145/3696410.3714729}.

\emph{Our study extends this literature through the first controlled evaluation of human detection performance against \textbf{adaptive, RL-based bots, not operating according to a static behavioral model}, testing 13 hypotheses on the roles of demographic traits, temporal learning, social interaction, and collective intelligence in shaping human detection capabilities.}

\section{Materials \& Methods}

\citet{10.1145/3696410.3714729} performed a controlled social media experiment involving 225 U.S. participants recruited on Amazon Mechanical Turk (MTurk). The current study uses the data collected in this experiment, only parts of which were used and released in \cite{10.1145/3696410.3714729}. \emph{We will release the additional data upon publication of this paper.}

\subsection{Participants}
 190 participants (84.4\%) were male and 35 (15.6\%) were female. The majority (71\%) were between 25 and 34 years old, with the remainder distributed across the 18--24, 35--44, and 45--54 age brackets. Most participants (97.2\%) reported holding a bachelor's or master's degree, and 87.4\% indicated either native English proficiency or full professional English competency. 

All participants reported familiarity with at least two major social media platforms (Facebook, Instagram, X/Twitter, TikTok, Reddit) and used social media for various purposes: 86.7\% for socializing, 77.8\% for news consumption, 78.1\% for professional networking, and 68\% for entertainment. 65\% expressed interest in all four topics used in the influence campaigns, while all participants indicated interest in at least two topics.

Of the 225 recruited participants, 86 actively engaged with the experimental platform, with 32 participating daily and 54 exhibiting sporadic engagement patterns. All experimental protocols were approved by Northwestern University's Institutional Review Board,  
and informed consent was obtained from all participants.

\subsection{Materials}

Because all major social platforms ban the creation of bot accounts and the running of CSIOs, we used the DartPost platform \cite{liu2021using} that was explicitly designed to mirror real-world social platforms for the purpose of running IRB-approved experiments in a legal and ethical manner. Following \cite{10.1145/3696410.3714729}, we used 305 accounts (225 humans, 80 bots) and ran four concurrent CSIOs, each controlling 20 bots through an RL\_CSIO agent. The campaigns promoted positive sentiment toward: (1) U.S. government COVID-19 response adequacy, (2) implementing a 2\% wealth tax on assets exceeding \$50 million, (3) ``Medicare for all who want it'' policy, and (4) GMO food safety. %

\noindent \textbf{RL\_CSIO Agent Architecture.}
The RL\_CSIO agent utilized is from \cite{10.1145/3696410.3714729} and uses
an actor-critic RL architecture. It models the social influence task as a Markov Decision Process (MDP) where the agent controls its bots to balance two competing objectives: maximizing influence spread while minimizing detection by an automated bot detection algorithm.

The agent architecture has three key components: (1) a Graph Convolutional Network (GCN) encoder that processed the social network state, including both graph topology and user attributes such as follower counts, posting behavior, and sentiment; (2) an actor network that generated probability distributions over possible actions for each controlled bot account; and (3) a critic network that estimated the value function for state-action pairs to guide policy optimization.

The state representation captures the dynamic social network at each timestep, encoding account attributes (role, centrality, polarity, blocked status, activity level) and network relationships. The action space includes posting content with specified sentiment polarity, following other accounts, liking posts, remaining inactive, and activating dormant accounts from the agent's reserve pool.

\noindent\textbf{Reward Structure and Learning Dynamics}
The RL\_CSIO agent \cite{10.1145/3696410.3714729} optimizes a multi-component reward function that captures the dual objectives of influence operations. The reward structure includes: (1) activation rewards for gaining new human followers; (2) infection rewards for influencing neighbors' sentiment toward the target topic; (3) termination rewards for achieving complete sentiment conversion; and (4) penalties when accounts were suspended by the bot detector.

This reward design encourages agents to develop sophisticated strategies that avoid detection while achieving influence. Agents learn to vary posting frequencies, adjust content sentiment, strategically time interactions, and manage account activation patterns to evade both AI and human detection mechanisms.

\noindent\textbf{Training and Deployment}
As in \cite{10.1145/3696410.3714729}, prior to deployment, RL\_CSIO agents underwent extensive training in synthetic environments populated with simulated accounts. The training environment replicated realistic social media dynamics, with simulated users exhibiting probabilistic behaviors for content creation, engagement, and inactivity. Simulated accounts updated their sentiment polarities through exponentially weighted moving averages based on content exposure, modeling gradual opinion shifts observed in real social influence processes.

During deployment, each CSIO campaign began with five active bot accounts, plus access to 15 additional accounts for strategic activation when suspensions of their active accounts occurred following detection. The social network was initialized with realistic follower distributions (3--6 connections per account on average) mimicking real-world social media topologies.

Following \cite{10.1145/3696410.3714729}, we used the bot detector proposed in \cite{9280525}, which employs a random forest classifier that periodically scanned all 305 accounts using behavioral features (e.g., posting frequency, follower growth patterns, and network characteristics). 

\subsection{Procedure}

Participants engaged with the DartPost platform~\cite{liu2021using} for at least 30 minutes daily over five consecutive days. DartPost replicated key functionalities of mainstream social media, including posting, liking, following, and content discovery features (e.g., a homepage with trends and most viral content).

Daily data collection also included comprehensive surveys administered at the end of each experiment day. These surveys captured: (1) participants' stance toward each campaign topic using 5-point Likert scales,  and (2) reports of accounts suspected to be bots. To understand whether participants were able to enhance their bot detection abilities over time in a natural setting without ground truth feedback, they were not told if the accounts they reported were really bots or not. To ensure data quality, surveys included attention checks requiring participants to correctly identify specific platform features and recall basic experimental parameters.

Additional behavioral data was automatically collected throughout each session, including all platform interactions (posts, likes, follows), content engagement patterns, and temporal activity distributions. Platform-generated metrics tracked account suspension events, interaction networks, and content propagation patterns.

Participants received daily email reminders to complete their tasks, with monetary incentives tied to task completion and bonus rewards for high engagement levels and accurate bot detection performance. Only participants who completed the daily requirements received compensation, though re-engagement in subsequent days was permitted regardless of previous participation gaps.


\section{Results}

\subsection{Demographic Characteristics}

To examine how individual characteristics influence bot detection capabilities, we analyzed performance differences across various demographic factors. Table~\ref{tab:demographics} presents F1-scores for the bot class across demographic groups, with statistical comparisons conducted using permutation testing \cite{good2013permutation,ernst2004permutation}. \emph{All $p$-values reported in this paper include adjustments for multiple hypothesis testing using the Benjamini–Hochberg false discovery rate (FDR) correction. Statistical significance was assessed using conventional thresholds: *$p < 0.05$, **$p < 0.01$, ***$p < 0.001$.
}

\begin{hyp} \rule{0.81\columnwidth}{0.5pt}
Female and male participants have similar bot detection performance.

\noindent\rule{\columnwidth}{0.5pt} 
\end{hyp}

Male participants achieved an F1-score of 0.454 ($N=65$) while female participants scored 0.435 ($N=21$), showing no statistically significant difference in bot detection abilities (FDR-corrected $p=0.902$). The performance gap of 0.019 between groups is not statistically significant, thus confirming our hypothesis.

\begin{hyp}     \rule{0.81\columnwidth}{0.5pt}
Younger participants (under 35 years old) are better at detecting bots compared to older participants (35 years and above).

\noindent\rule{\columnwidth}{0.5pt} 
\end{hyp}

We split participants based on whether they were younger or older than 35 years because people under 35 are considered to be "digital natives", i.e., the first generation to grow up entirely immersed in digital environments, social media, and online interactions from early childhood\footnote{\url{https://www.techtarget.com/whatis/definition/digital-native}}.
Contrary to expectations, older participants ($\geq$ 35 years) significantly outperformed younger participants ($<$ 35 years) in bot detection, achieving F1-scores of 0.570 versus 0.344 respectively (FDR-corrected $p=0.045$). The older group ($N=25$) demonstrated a substantial 0.226-point advantage over the younger group ($N=61$). Not only is our hypothesis rejected, but in fact, \emph{people over 35 years of age performed better than those below 35 at detecting bots.}

\begin{table}[t]
\centering
\caption{Bot Detection Performance by Demographics. F1-scores represent performance on the bot class. Statistical comparisons are made between different groups using permutation testing.  }
\label{tab:demographics}
\begin{tabular}{cccccc}
\toprule
\textbf{Demographic} & \textbf{Group} & \textbf{N} & \multicolumn{2}{c}{\textbf{F1}} & \textbf{FDR-corrected} \\ \cmidrule(lr){4-5}
\textbf{Factor} & & & $\mu$ & $\sigma$ & \textbf{p-value} \\

\midrule
\multirow{2}{*}{Gender} & Male   & 65  & 0.454 & 0.030 & \multirow{2}{*}{$0.902$} \\
                        & Female & 21  & 0.435 & 0.047 & \\
\midrule
\multirow{2}{*}{Age} & $< 35$    & 61 & 0.344 & 0.044 & \multirow{2}{*}{$0.045^*$} \\
                     & $\geq 35$ & 25 & 0.570 & 0.032 & \\
\midrule
\multirow{2}{*}{\shortstack{English\\Proficiency}} 
    & Native                     & 49 & 0.516 & 0.028 & \multirow{2}{*}{$0.009^{**}$} \\
    & Non-native                 & 37 & 0.153 & 0.025 & \\
\midrule
\multirow{2}{*}{Education} & Master's+   & 72 & 0.068 & 0.023 & \multirow{2}{*}{$0.028^*$} \\
                           & Bachelor's- & 14 & 0.542 & 0.021 & \\
\midrule
\multirow{2}{*}{\shortstack{Twitter\\Usage}} 
                            & User       & 57 & 0.496 & 0.032 & \multirow{2}{*}{$0.646$} \\
                            & Non-user   & 29 & 0.350 & 0.043 & \\
\midrule
\multirow{2}{*}{\shortstack{Reddit\\Usage}} 
                            & User       & 30 & 0.277 & 0.061 & \multirow{2}{*}{$0.208$} \\
                            & Non-user   & 56 & 0.538 & 0.029 & \\
\bottomrule
\end{tabular}%

\end{table}

\begin{hyp}     \rule{0.81\columnwidth}{0.5pt}
Native speakers have higher bot detection capabilities than non-native English speakers.

\noindent\rule{\columnwidth}{0.5pt} 
\end{hyp}

Native English speakers substantially outperformed non-native speakers, achieving F1-scores of 0.516 versus 0.153 respectively (FDR-corrected $p=0.009$). This represents a striking 0.363-point performance gap, with native speakers ($N=49$) showing more than three times the detection capability of non-native speakers ($N=37$). Therefore,  our hypothesis is confirmed - native speakers exhibit significantly better bot detection capabilities than non-native English speakers. This substantial difference may reflect deeper familiarity with the specific topics promoted by the bots rather than purely linguistic advantages, i.e., native English speakers located in the U.S. may possess greater contextual knowledge about the influence campaigns discussed in the experiment (e.g., U.S. government's COVID-19 response adequacy, or "Medicare for all who want it" policies).

\begin{hyp}     \rule{0.81\columnwidth}{0.5pt}
Participants with a higher educational background (master's degree or above) are better at detecting bots than those with lower educational levels. 

\noindent\rule{\columnwidth}{0.5pt} 
\end{hyp}

Participants with lower educational levels (bachelor's degree or below) dramatically outperformed those with master's degrees or higher, achieving F1-scores of 0.542 versus 0.068 respectively (FDR-corrected $p=0.028$). \emph{Despite the smaller sample size ($N=14$ vs $N=72$), the lower-education group showed nearly eight times better detection performance. } Our hypothesis is thus rejected. {  This result may be because highly educated subjects had preconceived notions of what bots should look like (e.g., repetitive posting patterns, poor grammar),  leading them to search for stereotypical indicators that RL-based bots quickly learned to avoid. }


\begin{hyp}     \rule{0.81\columnwidth}{0.5pt}
Participants who use social media platforms are better at detecting bots compared to those who do not.

\noindent\rule{\columnwidth}{0.5pt} 
\end{hyp}

{ 
Since bots primarily operate within social platforms, we expect that users who frequently use these platforms would develop intuitions about authentic user behavior, content patterns, and interaction dynamics that could translate into better bot detection performance. We examined two platforms: Twitter (now X) and Reddit, given their prominence in bot-related discussions  \cite{xu2025social, ma-lalor-2020-empirical, Rizoiu_Graham_Zhang_Zhang_Ackland_Xie_2018, Hegelich_Janetzko_2021} and DartPost's similarity to Twitter's interface. For each platform, "non-users" refers to participants who reported not using that specific platform, rather than non-users of social media generally.
}

Our results show mixed, non-significant patterns across platforms. Twitter users ($N=57$) achieved a higher F1-score of 0.496 compared to 0.350 for non-users ($N=29$), but the 0.146-point difference was not statistically significant (FDR-corrected $p=0.646$). Conversely, participants who do not use Reddit ($N=56$) outperformed Reddit users ($N=30$) with F1-scores of 0.538 versus 0.277, representing a 0.261-point advantage, though this difference was also not statistically significant (FDR-corrected $p=0.208$). 

{  To further disentangle platform-specific effects, we compared Twitter-only users (those reporting Twitter but not Reddit usage) against all other participants and found no significant difference (FDR-corrected $p=0.342$). An insufficient number of Reddit-only users ($N=5$) prevented symmetric analysis. 

Overall, our hypothesis is rejected: social media platform usage does not systematically improve bot detection performance. This result is in line with prior work~\cite{kenny2024duped} showing that greater social media experience does not guarantee improved accuracy and may foster overconfidence.
}





\subsection{Temporal Learning Effects}

\begin{hyp}     \rule{0.81\columnwidth}{0.5pt}\label{hyp:temporal}
Participants' bot detection performance improves progressively throughout the five-day experiment. 

\noindent\rule{\columnwidth}{0.5pt} 
\end{hyp}

{ 
RL-based bots in our experiment were trained to evade an automatic detector rather than human judgment. Therefore, we study if humans implicitly learnt bots' behavioral patterns over time, even without explicit feedback on their reports.

We analyzed performance over five days using two methods. \emph{Day-specific} analysis evaluated detection accuracy separately for each day, classifying an account as a bot if it was reported at least once that day. The \emph{cumulative} approach considered an account as a bot if it was reported on any day up to and including the current one.

}

\begin{figure}[t]
     \centering

     \subfloat[][]{\includegraphics[width=.4\linewidth]{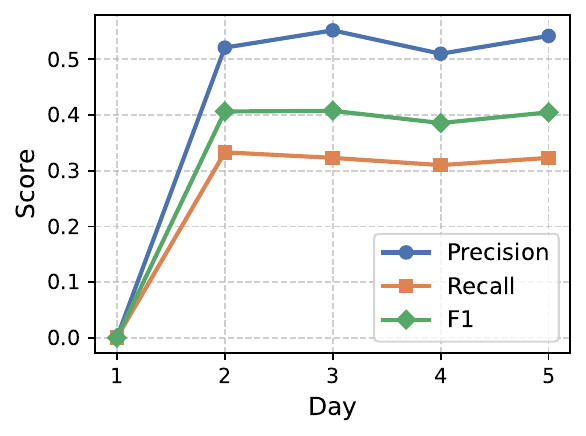}\label{fig:day-specific}}
     \subfloat[][]{\includegraphics[width=.4\linewidth]{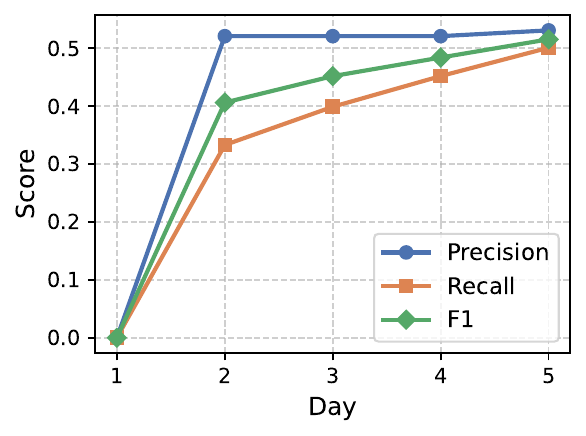}\label{fig:cumulative}}
     
     \caption{Human bot detection performance across the five-day experiment. (a) Day-specific: accounts classified as bots if reported on that specific day. (b) Cumulative: accounts classified as bots if reported on any day up to and including the current day.}
    
     \label{fig:time-based}
\end{figure}

Results indicate no evidence that human participants learned over time for both the day-specific and cumulative approaches. There were only five incorrect reports on day 1, yielding zero precision, recall, and F1-scores on the bot class. From day 2 onward, performance was stable, with minimal day-to-day variation across all metrics (see Figure \ref{fig:time-based} for visual trends). Linear regression models using \emph{day} as the predictor of F1 score confirmed no improvement: for the \emph{day-specific} analysis, $\beta = 3.10 \times 10^{-4}$ (FDR-corrected $p = 0.987$), and for the \emph{cumulative} analysis, $\beta = 5.44 \times 10^{-2}$ ($p = 0.379$), both statistically non-significant.  We therefore reject our hypothesis: \emph{Human participants exhibited no measurable learning or adaptation, maintaining consistent detection performance throughout the experiment.}

\subsection{Social Media Activity}

To examine the relationship between human social media activity and their bot detection ability, we define two complementary metrics: (i) Bot Engagement Ratio ($BER$) measures the proportion of a user's outgoing interactions directed toward bot accounts, calculated separately for likes ($BER^{\text{like}}$) and follows ($BER^{\text{follow}}$). (ii) Bot Exposure Ratio ($BXR$) is the proportion of incoming interactions a user receives from bots, again measured for both likes ($BXR^{\text{like}}$) and follows ($BXR^{\text{follow}}$). These metrics capture distinct aspects of human-bot interaction: BER reflects users' voluntary engagement with bot-generated content, while BXR represents exposure to bot activity regardless of user intent.

\begin{figure}
     \centering
     \subfloat[][]
     {\includegraphics[width=.24\textwidth]{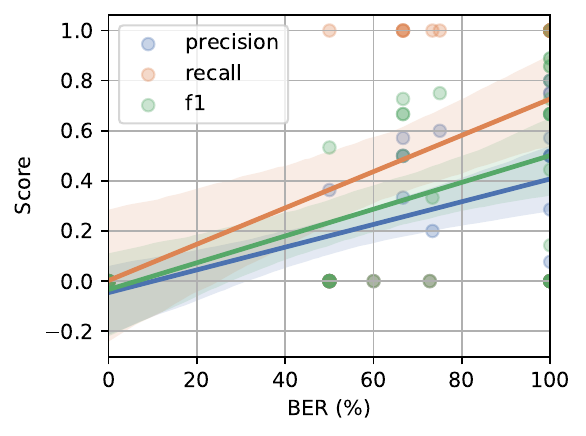}\label{fig:BER_follows}}
     \subfloat[][]{\includegraphics[width=.24\textwidth]{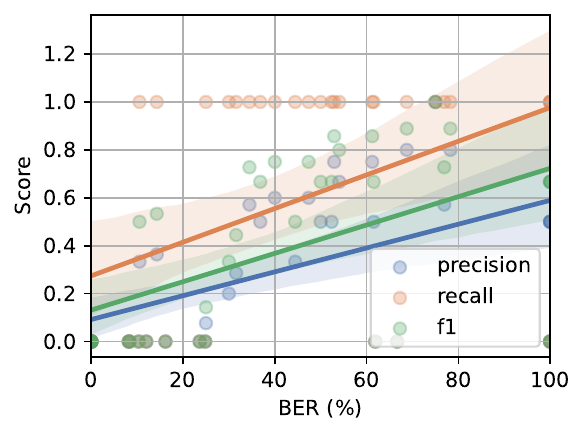}\label{fig:BER_likes}}
     \subfloat[][]
     {\includegraphics[width=.24\textwidth]{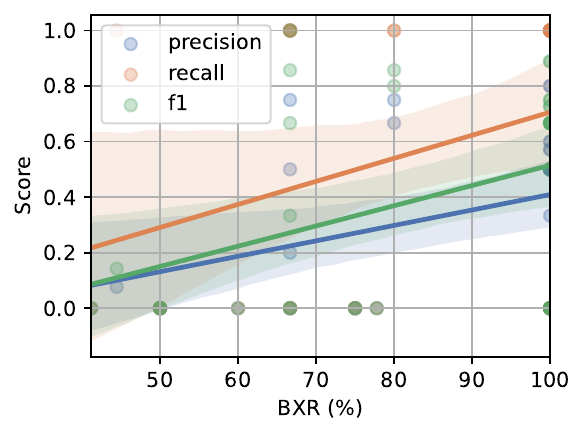}\label{fig:BXR_follows}}
     \subfloat[][]
     {\includegraphics[width=.24\textwidth]{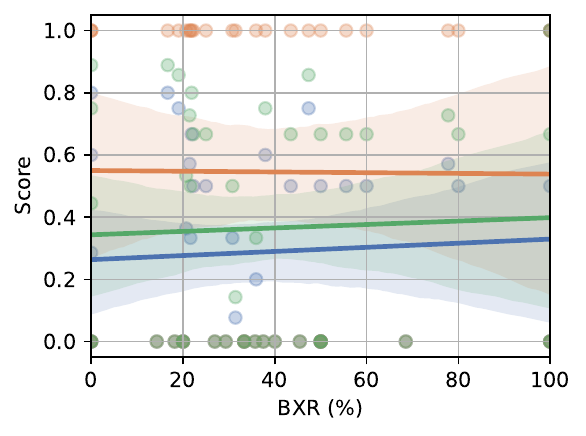}\label{fig:BXR_likes}}
     \caption{Relationship between bot interaction patterns and detection performance. (a-b) Bot Engagement Ratio (BER): proportion of user's outgoing interactions (follows-likes) directed toward bots. (c-d) Bot Exposure Ratio (BXR): proportion of incoming interactions (follows-likes) received from bots.}

     \label{fig:social_media_activity}
\end{figure}

\begin{hyp} 
\rule{0.81\columnwidth}{0.5pt} 

Users who engage more frequently with bots have a higher bot detection performance compared to participants with lower engagement levels.

\noindent\rule{\columnwidth}{0.5pt}
\end{hyp}

Figures \ref{fig:BER_follows} and \ref{fig:BER_likes} present detection performance metrics (precision, recall, F1-score on the bot class) as a function of $BER^{\text{follow}}$ and $BER^{\text{like}}$, respectively. We observe positive trends across all metrics. To validate these results, we also fitted linear regression models with detection performance as the dependent variable and $BER^{\text{follow}}$ and $BER^{\text{like}}$ as predictors. For $BER^{\text{follow}}$, we find $\beta=0.004$ (FDR-corrected $p=0.003$) for precision, 
$\beta=0.007$ (FDR-corrected $p=0.003$) for recall, and 
$\beta=0.005$ (FDR-corrected $p=0.003$) for F1-score. Similarly, for $BER^{\text{like}}$, we observe 
$\beta=0.005$ (FDR-corrected $p=1.7 \times 10^{-4}$) for precision, 
$\beta=0.007$ (FDR-corrected $p=0.001$)   for recall, and 
$\beta=0.006$ (FDR-corrected $p=7.1 \times 10^{-5}$) for F1-score. Table \ref{tab:app_social_activity} in Appendix \ref{app:social_media_activity} shows  full regression statistics. Overall, these results confirm our hypothesis:
\emph{users who engage more frequently with bot accounts develop enhanced detection capabilities.}

\begin{hyp}     \rule{0.81\columnwidth}{0.5pt}
Users who receive a higher volume of interactions initiated by bot accounts are better at detecting bots than those who receive fewer such bot interactions.

\noindent\rule{\columnwidth}{0.5pt} 
\end{hyp}

Figures \ref{fig:BXR_follows} and \ref{fig:BXR_likes} show detection performance metrics (precision, recall, F1-score) as a function of bot exposure ratios, $BXR^{\text{follow}}$ and $BXR^{\text{like}}$, respectively. We fitted analogous linear regression models using $BXR^{\text{like}}$ and $BXR^{\text{follow}}$ as predictors.

The patterns diverge substantially between interaction types. For $BXR^{\text{like}}$, we observe a positive relationship similar to the engagement metrics: $\beta=0.005$ (FDR-corrected $p=0.033$) for precision, $\beta=0.008$ (FDR-corrected $p=0.044$) for recall, and 
$\beta=0.007$ (FDR-corrected $p=0.017$) for F1-score. This aligns with evidence from the original experiment~\cite{10.1145/3696410.3714729} where increased views of bot content improved human detection capabilities. In contrast, $BXR^{\text{follow}}$ shows no significant relationship with detection performance (FDR-corrected $p>0.70$). {  This difference likely depends on the nature of each interaction type. Each time that a bot likes a user's post, the user receives a notification in their Dartpost's feed, and may inspect the bot's profile, providing direct exposure to the bot's behavioral patterns. In contrast, when a bot follows a user, this generates only a one-time notification and does not entail ongoing exposure to the bot's content or activity. 

These results provide partial support for our hypothesis: \emph{users receiving observable interactions from bots (likes) show improved detection performance, while passive follower relationships which do not translate into behavioral exposure show no effect.}

Taken together, the results from both hypotheses reveal significant differences between active engagement ($BER$) by users and passive exposure ($BXR$). The consistent positive effects of $BER$ across both likes and follows, compared with the weak and inconsistent effects of $BXR$, suggest that active engagement with bot content drives improvement in humans' detection ability. Users who voluntarily interact with bots may develop mental models of bot behavior patterns through direct experience which may help them better detect bots in the future. Passive exposure, by contrast, improves detection only when it involves observable interactions (likes) that prompt users to examine bot accounts, rather than mere follower relationships that provide no behavioral information.
}

\begin{figure}[t]
     \centering

    \includegraphics[width=.6\textwidth]{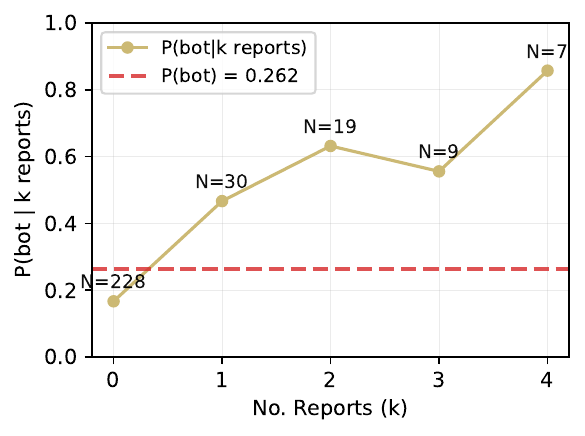}
     
    \caption{ Conditional probability of bot given it was reported ${k}$ times, ${P(bot|k)}$. The red dashed line shows the overall proportion of bots in the original experiment (0.262).}

     \label{fig:performanec_by_n_reports}
\end{figure}

\subsection{Collective Intelligence}

\begin{hyp}     \rule{0.81\columnwidth}{0.5pt} \label{hyp:mtimes}
Accounts reported as bots multiple times have a higher probability of actually being bots compared to accounts reported by single participants.

\noindent\rule{\columnwidth}{0.5pt} 
\end{hyp}

Multiple participants, even across different days, could report accounts as bots. Figure \ref{fig:performanec_by_n_reports} shows the conditional probability that an account is a bot given it was reported $k$ times:
\begin{equation*}
    P(bot|k)=\frac{\text{No. of actual bot accounts reported $k$ times }}{\text{No. of accounts reported $k$ times}}
\end{equation*}

In line with \cite{10.1145/3696410.3714729}, the baseline probability of an account being a bot is $0.262$ (80 bots out of 305 accounts). We observe a clear positive relationship between report frequency and bot probability. Accounts reported once ($N=30$) show $P(bot|k=1) = 0.467$, nearly double the baseline rate. This probability increases further to $0.632$ for accounts reported twice ($N=19$), and to $0.857$ for accounts reported four times ($N=7$). We omit results for larger $k$ as they are extremely sparse, i.e., $N \leq 3$ for any $k \in [5,11]$.

Interestingly, unreported accounts ($N=228$) exhibit $P(bot|k=0) = 0.167$, substantially below the baseline ($0.262$). This indicates that unreported accounts are less likely to be bots than the average account. However, the fact that 75\% (228 out of 305) of accounts—including 38 actual bots—were unreported exposes a major coverage limitation: collective detection, while accurate where it occurs, operates over a narrow subset of the population, potentially missing bots.

These results provide strong support for our hypothesis: \emph{report frequency is a reliable, but imperfect, indicator of bot likelihood.} However, they also highlight a fundamental trade-off: aggregated human reporting exhibits valuable collective intelligence but suffers from severe coverage constraints that may limit its systemic effectiveness. 


\begin{hyp}     \rule{0.81\columnwidth}{0.5pt} \label{hyp:f1-weighted}
Aggregating reports weighted by reporters' F1-scores achieves better bot detection performance than classifying bots based solely on whether they exceed a fixed number of reports.

\noindent\rule{\columnwidth}{0.5pt} 
\end{hyp}

We test whether incorporating the quality of each reporter further enhances detection accuracy. While multiple reports generally increase confidence, not all participants display the same detection ability. To address this heterogeneity, we compare two aggregation strategies that differ in how they combine human judgments.

The \emph{Count-based strategy} classifies an account as a bot if there are at least $k$ independent reports that it is a bot:
\begin{equation*}
a \text{ is bot} \iff |R(a)| \geq k
\end{equation*}
where $R(a)$ denotes the set of reports for account $a$, and $k \in \{1,2,3,4,5\}$ is a reporting threshold. This approach assumes all reports are equally informative, regardless of reporter skill.

The \emph{Quality-weighted strategy} instead accounts for reporter reliability by weighting each report by the reporter's F1-score:
\begin{equation*}
s(a) = \sum_{i \in R(a)} F1_i
\end{equation*}
An account is deemed a bot if $s(a) \geq \tau = 0.533$. This threshold (0.533) represents the mean F1-score across all reporters, and corresponds to the evidence contributed by one average-performing reporter\footnote{This setting simulates a realistic scenario in which platforms maintain reporter reliability scores via periodic audits. Because participants received no feedback and showed no significant learning effects (Hypothesis~\ref{hyp:temporal}), their end-of-experiment performance serves as a reasonable proxy for capabilities throughout this study.}. 

Both strategies operate independently of ground-truth labels and are equally affected by the coverage limitation observed earlier (Hypothesis~\ref{hyp:mtimes}), whereby many accounts remain unreported. 


{ 
Table~\ref{tab:aggregation_strategies} summarizes performance. As expected, increasing $k$ in the count-based strategy reduces recall by overlooking true bots. However, precision does not improve consistently, peaking at $k=3$ (0.666) and then declining for higher $k$ values, indicating that frequent reports from uninformed users can reduce accuracy. The highest F1-score (0.549) occurs at $k=1$, suggesting that report volume alone is an unreliable indicator of detection quality.

Comparing the best \emph{Count-based} strategy ($k=1$) to the \emph{Quality-weighted} approach reveals an advantage for the latter: precision increases by 14.4\% (0.575 $\rightarrow$ 0.672), recall decreases slightly by -2.3\%, leading to a 5.67\% relative F1 improvement (0.549 $\rightarrow$ 0.582). The two methods agree on 92.7\% of classifications, diverging on 12 accounts—all correctly resolved by the \emph{Quality-weighted} strategy. We find that this difference is statistically significant. A Chi-square test yields $\chi^2 = 117.2$, FDR-corrected $p=1.99 \times 10^{-27}$, and McNemar’s test on disagreements yields $\chi^2 = 1.0$, FDR-corrected for multiple hypothesis testing $p = 0.006$. 

\emph{Overall, these findings provide strong support for Hypothesis~\ref{hyp:f1-weighted}, highlighting that incorporating reporter reliability can enhance collective human detection performance.}

}
\begin{table}[t]
\centering
\caption{Performance comparison of two aggregation strategies in terms precision, recall, F1-score on the bot class along with overall accuracy. Count-based: flag account as bot if reported at least $k$ times (treating all reports equally). Quality-weighted: give more weight to reports from human reporters who performed better (in terms of F1-score) at bot detection, then flag accounts exceeding a weighted score threshold $\tau=0.533$ (mean reporter F1). }

\label{tab:aggregation_strategies}

\begin{tabular}{@{}lccccc@{}}
\toprule
\textbf{Strategy} & \textbf{Threshold} & \textbf{Precision} & \textbf{Recall} & \textbf{F1} & \textbf{Accuracy} \\
\midrule
\multicolumn{6}{l}{\textit{Count-based: All reports weighted equally}} \\
\midrule
Count-based & $k = 1$ & $0.575$ & $0.525$ & $0.549$ & $0.584$ \\
Count-based & $k = 2$ & $0.589$ & $0.413$ & $0.485$ & $0.578$ \\
Count-based & $k = 3$ & $0.666$ & $0.325$ & $0.436$ & $0.596$ \\
Count-based & $k = 4$ & $0.647$ & $0.275$ & $0.385$ & $0.578$ \\
Count-based & $k = 5$ & $0.608$ & $0.175$ & $0.272$ & $0.548$ \\
\midrule
\multicolumn{6}{l}{\textit{Quality-weighted vs Count-based}} \\
\midrule
Count-based & $k = 1$ & $0.575$ & $0.525$ & $0.549$ &  $0.584$ \\
Quality-weighted & \begin{tabular}[c]{@{}c@{}}Mean F1\\(0.533)\end{tabular} & $0.672$ & $0.513$ & $0.582$ & $0.645$ \\

\cmidrule{1-6}
\textit{Difference} & & \textit{+14.4\%} & \textit{-2.30\%} & \textit{+5.67\%} & \textit{+9.46\%}\\

\bottomrule
\end{tabular}

\end{table}



\subsection{Human Ensemble vs. ML Bot Detectors}

We evaluated the performance of human ensembles (``crowds'') against several SOTA AI detectors. Individual human predictions were aggregated using the \emph{quality-weighted} strategy (cf. Hypothesis~\ref{hyp:f1-weighted}), where each report is weighted by the reporter’s individual F1-score. 

We compared the human ensemble with three recent bot detectors: \emph{(i)} \emph{RFS} \cite{9280525} and \emph{(ii)} \emph{BotBuster} \cite{ng2023botbuster} employ hand-crafted features based on behavioral patterns and metadata (e.g., follower counts, posting frequency). 
\emph{(iii)} \emph{LLM-based Detector} \cite{feng-etal-2024-bot} utilizes a prompt-based approach with in-context examples for bot identification. We instantiated this baseline with four LLMs: \texttt{llama3.1:8b}, \texttt{llama3.1:70b}, \texttt{mistral:7b}, \texttt{mistral:24b}.

\begin{hyp}     \rule{0.81\columnwidth}{0.5pt}
\label{hyp:h_better_m}
Human ensembles are better at detecting bots than AI bot detectors.

\noindent\rule{\columnwidth}{0.5pt} 
\end{hyp}

All AI detectors were pretrained on standard benchmarks—Twibot-20~\cite{10.1145/3459637.3482019}, Cresci-2017~\cite{10.1145/3041021.3055135}, and Caverlee-2011~\cite{Lee_Eoff_Caverlee_2021}—and deployed on our experimental dataset\footnote{Fine-tuning on our experimental data was not performed to maintain realism as, in the real-world, we would not be able to fine-tune models on an actual bot campaign before it started.}. This setup emulates realistic operational conditions in which detectors face novel, unseen influence campaigns. See Appendix~\ref{app:implementation} for benchmark performance\footnote{All detectors exhibited lower performance on our experimental data compared to their original benchmark test sets, reflecting the challenges of out-of-distribution (OOD) evaluation \cite{10.1145/3701716.3715495,yang2020scalable} and detecting adaptive bots powered by RL. Although methods like BotBuster emphasize improved generalization to unseen data, addressing OOD robustness is beyond the scope and contributions of this study.} We assess human performance over all 305 accounts, treating the human ``crowd'' as a full ensemble-based detection system in which unreported accounts are implicitly considered human.

\begin{table}[t]
\centering
\caption{Human Ensemble vs. AI Bot Detection Performance: Precision, Recall, and F1-score on the bot class, along with overall Accuracy. For all detectors except the LLM-based baseline, "Training Dataset" refers to the dataset used for training. For the LLM-based baseline, the column lists the LLM under analysis. Bold indicates the best performance, and underline indicates the first runner-up.}
\label{tab:human_vs_algo}
\begin{tabular}{rccccc}
\toprule
\textbf{Method} & \textbf{ Training Dataset} & \textbf{Precision} & \textbf{Recall} & \textbf{F1} & \textbf{Accuracy} \\
\midrule
\multicolumn{5}{c}{\textit{Human Ensemble Detection}} \\
Humans & N/A & $\mathbf{0.672}$ & 0.513 & 0.582 & $\mathbf{0.645}$ \\
\midrule
\multicolumn{5}{c}{\textit{AI Detection}} \\
\multirow{3}{*}{BotBuster \cite{ng2023botbuster}}  & Twibot-20   & \underline{0.564} & 0.600 & 0.582  & \underline{0.558} \\
                            & Cresci-2017 & 0.539 & 0.425 & 0.476 & 0.519 \\
                            & Caverlee-2011 & 0.437 & 0.388 & 0.411 & 0.429 \\
\midrule
\multirow{3}{*}{RFS \cite{9280525}}        & Twibot-20   & 0.491 & 0.650 & 0.559 & 0.474\\
                            & Cresci-2017 & 0.400 & 0.400 & 0.400 & 0.385\\
                            & Caverlee-2011 & 0.513 & $\mathbf{1.000}$ & $\mathbf{0.678}$ & 0.513\\
\midrule

\multirow{4}{*}{LLM-based \cite{feng-etal-2024-bot}}  

& \texttt{llama 3.1:8b} & 0.501 & 0.448 & 0.474 & 0.530 \\
& \texttt{llama 3.1:70b} & 0.494 & 0.475 & 0.484 & 0.481\\
& \texttt{mistral:7b} & 0.350 & 0.087 & 0.140 & 0.482 \\
& \texttt{mistral:24b} & 0.513 & \underline{0.950} & \underline{0.667} & 0.513 \\

\bottomrule
\end{tabular}%

\end{table}

\begin{figure}[t]
     \centering
     \subfloat[][]
     {\includegraphics[width=.4\textwidth]{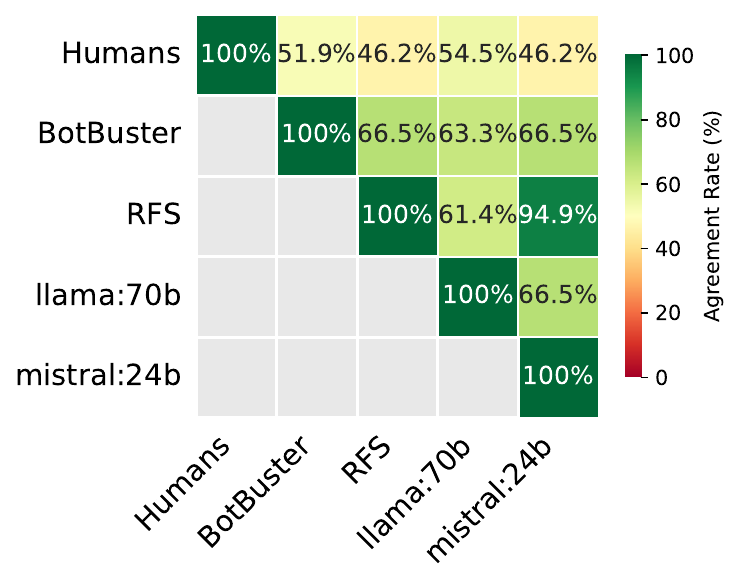}\label{fig:h_a_agreement_rate}}
     \subfloat[][]{\includegraphics[width=.4\textwidth]{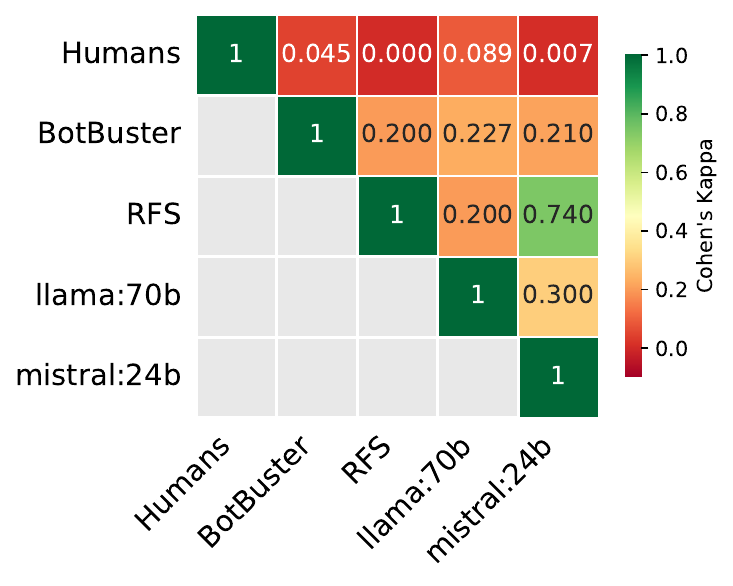}\label{fig:h_a_cohen}}
     
    \caption{Human vs AI Comparison: Pairwise Agreement Rate (a) and Cohen's $\kappa$ (b) for humans and the best performing detectors, i.e., BotBuster (trained on Twibot-20), RFS (trained on Caverlee-2011), 
    LLM-based (\texttt{llama:70b}), and LLM-based (\texttt{mistral-24b}).}

     \label{fig:h_a_agreement}
\end{figure}

{ 
Table \ref{tab:human_vs_algo} presents performance in terms of precision, recall, F1-score on the bot class, and overall accuracy. Human ensembles achieved the highest precision on the bot class (0.672) and overall accuracy (0.645), while AI approaches generally attained higher recall. 

However, these patterns reflect fundamentally different failure modes rather than calibration differences.

One AI detector (RFS trained on Caverlee-2011) collapses by predicting all accounts as bots. Indeed, it achieved perfect recall (1.000) but near-chance precision (0.513). Among the others, a consistent recall-over-precision bias emerges: RFS on Twibot-20 (precision 0.491, recall 0.650), \texttt{mistral:24b} (precision 0.513, recall 0.950), and BotBuster on Twibot-20 (precision 0.564, recall 0.600). 

Humans show the inverse pattern, higher precision than recall, but for a distinct reason: incomplete coverage rather than conservative classification. Of the 305 accounts, 228 (including 38 bots) were never reported by any participant, thus reducing recall. Importantly, this does not indicate that humans are more cautious classifiers; rather, they only evaluate a subset of accounts, and within that subset, their precision remains 0.672. The coverage gap, not classification threshold, drives the precision-recall asymmetry.



To understand whether humans and AI algorithms identify the same bots, we analyzed pairwise agreement across all detection methods and humans. Figure \ref{fig:h_a_agreement} presents pairwise agreement in terms of raw agreement rate and Cohen's $\kappa$ values between humans and the best-performing AI models. We find minimal agreement: humans and AI algorithms achieved agreement rates ranging from 46.2\% (RFS) to 54.5\% (\texttt{llama:70b}), with $\kappa$ values close to zero ($\kappa \in [0.0, 0.089]$). Similarly, AI methods have little mutual consistency (e.g., BotBuster vs. \texttt{llama:70b}: $\kappa = 0.20$). The only exception is the RFS-\texttt{mistral:24b} pair where agreement rate is 94.9\% and $\kappa=0.74$, and it depends on the fact that the two detectors tend to classify every account as a bot (bot recall $> 0.9$ in Table \ref{tab:human_vs_algo}).  

This lack of concordance demonstrates that humans and AI algorithms do not merely differ in accuracy but also in the type of evidence they rely upon. Rather than detecting the same bots with varying success, they appear to target disjoint subsets of accounts using fundamentally different heuristics and behavioral indicators. \emph{Overall, these findings partially support our hypothesis: humans show superior precision but markedly lower recall and F1-scores than AI detectors.} In addition, the low agreement between humans and AI algorithms indicates complementarity rather than redundancy. 
}

{ 
\begin{hyp}     \rule{0.81\columnwidth}{0.5pt}
\label{hyp:agg_str}

Combining human reports with AI predictions improves bot detection performance compared to using either source alone.

\noindent\rule{\columnwidth}{0.5pt} 
\end{hyp}

We combine human reports and AI predictions using 7 strategies in two categories: model-only ensembles that combine AI detectors, and hybrid ensembles that additionally incorporate human reports.

\begin{itemize}
    \item \emph{Model-only ensemble}: We evaluate three standard ensemble methods: (i) \emph{Hard Voting}, which assigns the majority label across detectors; (ii) \emph{Soft Voting}, which averages predicted probabilities; and (iii) \emph{Late Fusion}, which computes a weighted average of probabilities. 
    \item \emph{Hybrid ensemble}: We evaluate four strategies that integrate human reports with AI predictions~\cite{10.1145/3626772.3657965}. (iv) \emph{Human First} applies soft voting to AI predictions to obtain an interim vote, then combines this interim vote with all individual human reports via hard voting. (v) \emph{Model First} aggregates human reports using the quality-weighted approach (cf. Hypothesis \ref{hyp:f1-weighted}), then combines this with AI predictions via soft voting. (vi) \emph{Meta Voting} enumerates all possible subgroups of voters (both human and AI), computes both hard and soft voting outcomes within each subgroup, and reaches a final decision by hard voting over all subgroup outcomes. (vii) \emph{Hybrid Late Fusion} extends the model-only variant by including human reports as an additional source, weighted by individual reporter F1-scores. 
    
\end{itemize}

These strategies require optimization on held-out data. As soft voting requires the selection of a probability threshold, we evaluated thresholds from 0.5 to 0.95 and selected 0.71 based on validation F1-score. As late fusion requires weight optimization, we randomly sampled 1,000 weight configurations (constrained to sum to 1) and retained the configuration with the highest validation F1-score. Hybrid strategies inherit these optimization procedures from their model-only counterparts.

To enable this optimization and ensure fair comparison across all strategies and individual models, we fine-tuned all detectors on our experimental data using 5-fold cross-validation\footnote{This setup resolves the out-of-distribution limitations observed in the previous hypothesis.}. We selected 5-fold over 10-fold cross-validation because it provides a more favorable training-to-test ratio (80\%-20\% vs. 90\%-10\%), ensuring sufficient training data for fine-tuning LLM-based detectors while maintaining stable evaluation. Specifically, we retrained BotBuster and RFS from scratch using Dartpost training data in each fold, while we fine-tuned \texttt{llama:70b} and \texttt{mistral:24b} using LoRA (Low-Rank Adaptation) for parameter-efficient updates.

Table~\ref{tab:cv-aggregation} presents results across all aggregation strategies.\footnote{We do not report AUC as it cannot be measured for aggregation strategies producing discrete classifications (e.g., Hard Voting, Meta Voting).}. Unsurprisingly, fine-tuning substantially improves individual detector performance compared to the pretrained-only setting (cf. Table~\ref{tab:human_vs_algo}): RFS achieves an F1-score of 0.723 (vs. 0.559 pretrained on Twibot-20), and LLM-based detection with \texttt{llama:70b} reaches 0.749 (vs. 0.484 pretrained). Human ensemble performance remains unchanged at 0.582 F1, as no fine-tuning applies.

\emph{Model-only ensembles} improve over individual detectors. Among these methods, \emph{Late Fusion} achieves the best precision (0.831), accuracy (0.777), and F1-score (0.745), while \emph{Soft Voting} attains the highest recall (0.850). However, hybrid ensembles yield further gains across most metrics.

\emph{Hybrid Late Fusion} achieves the highest precision (0.871) and accuracy (0.795), improving +4.0 percentage points in precision and +1.8 points in accuracy over its model-only counterpart. It also achieves the second-best F1-score (0.761) among all methods. This gain stems from incorporating high-performing human reports without inheriting their coverage limitation: the optimized model weights maintain recall at 0.675 while human contributions boost precision.

\emph{Meta Voting} achieves the highest F1-score (0.801), outperforming the best individual detector (\texttt{llama:70b}, F1 = 0.749) by 5.2 percentage points and the best model-only ensemble (\emph{Late Fusion}, F1 = 0.745) by 5.6 points. Unlike late fusion, which maximizes precision, \emph{Meta Voting} achieves a more balanced precision-recall trade-off by leveraging agreement patterns across all possible voter combinations: when any subgroup of humans and detectors agree on an account, that agreement gets captured even if other sources disagree. In contrast, \emph{Human First} and \emph{Model First} require majority agreement, missing bots detected by only a few sources.

\emph{Overall, these findings support our hypothesis: combining human reports with AI predictions improves bot detection performance. } 
\clearpage

\begin{table}[t]
\centering
\caption{Aggregation strategies for bot detection. ``No Aggregation'' shows individual detector and human ensemble performance after fine-tuning. ``Models-only'' combines AI detectors. ``Hybrid'' integrates human reports with AI predictions. Bold indicates best performance per metric; underline indicates second-best performer.}
\label{tab:cv-aggregation}
\begin{tabular}{cccccc}
\toprule

 & \begin{tabular}{c} \textbf{Aggregation}  \\ \textbf{Method} \end{tabular}
 & \textbf{Precision} & \textbf{Recall} & \textbf{F1} & \textbf{Accuracy} \\
\midrule

\multirow{6}{*}{\begin{tabular}{c} No  \\ Aggregation\end{tabular} }

& Humans  & 0.672 & 0.513 & 0.582 & 0.645 \\ 
& BotBuster & 0.567 & 0.675 & 0.606 & 0.605 \\
& RFS  & 0.739 & 0.708 & 0.723 & 0.719 \\ 
& \begin{tabular}{c} LLM-based  \\ (\texttt{llama:70b}) \end{tabular} & 0.709 & 0.793 & 0.749 & 0.684 \\ 
& \begin{tabular}{c} LLM-based  \\ (\texttt{mistral:24b}) \end{tabular} & 0.688 & 0.766 & 0.703 & 0.619 \\
\midrule

\multirow{3}{*}{AI Models-only}

& Hard Voting & 0.677 & 0.811 & 0.738 & 0.723 \\
& Soft Voting & 0.660 & $\mathbf{0.850}$ & 0.740 & 0.717 \\
& Late Fusion & \underline{0.831} & 0.675 & 0.745 & \underline{0.777} \\
\midrule

\multirow{4}{*}{Hybrid}

& Human First  & 0.677 & \underline{0.812} & 0.738 & 0.723 \\
& Model First  & 0.723 & 0.750 & 0.736 & 0.741 \\
& Meta Voting & 0.741 & 0.746 & $\mathbf{0.801}$ & 0.761 \\ 
& Late Fusion  & $\mathbf{0.871}$ & 0.675 & \underline{0.761} & $\mathbf{0.795}$ \\

\bottomrule
\end{tabular}%

\end{table}

}

\begin{figure*}
    \centering
    \includegraphics[width=\textwidth]{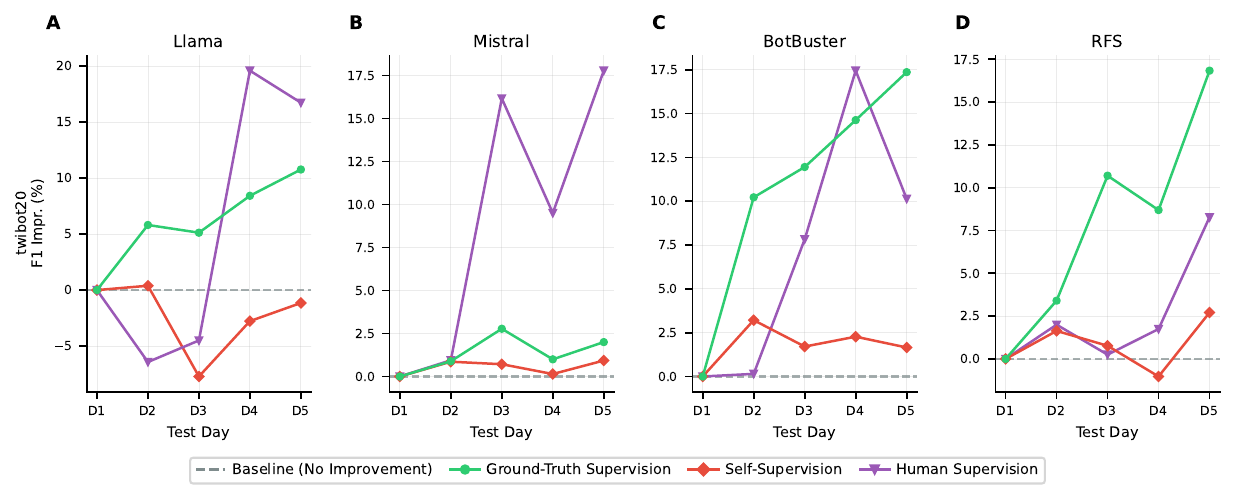}
    \caption{Relative F1-score improvement over baseline (no retraining) across experimental days for four detectors pretrained on Twibot-20. Each subplot shows one detector. \emph{Ground-Truth Supervision} retrains on correct predictions, \emph{Self-Supervision} retrains on high-confidence predictions (threshold 0.7), and \emph{Human Supervision} retrains on human reports.}
    \label{fig:rt_twibot}
\end{figure*}

{ 
\begin{hyp}     \rule{0.81\columnwidth}{0.5pt}
\label{hyp:rt_s}

Human reports provide effective supervision for incrementally retraining bot detectors.

\noindent\rule{\columnwidth}{0.5pt} 
\end{hyp}

We now investigate whether human reports can improve detectors at training time by serving as supervision for incremental retraining. We design a time-based retraining protocol that simulates incremental deployment. On each day $D$, we retrain each detector using its original benchmark data augmented with accounts discovered on days 1 through $D-1$, then evaluate on accounts on day $D$. This protocol mirrors realistic deployment where detectors are periodically updated with newly acquired data.

We compare three supervision strategies for selecting retraining instances:
\begin{itemize}
    \item \emph{Ground-Truth Supervision.} The detector is retrained on instances where its previous bot predictions were correct. This strategy represents an ideal scenario because it requires access to ground-truth labels.

    \item \emph{Self-Supervision.} The detector is retrained on instances it predicted with confidence exceeding a reasonable threshold. This strategy requires no access to ground-truth. We evaluated thresholds from 0.5 to 0.95 and selected 0.7, which achieved the best performance in terms of F1-score across all detectors. See Appendix \ref{app:ss_ret} for additional results. 

    \item \emph{Human Supervision.} The detector is retrained on instances reported by human participants aggregated using the \emph{quality-weighted} strategy. 
\end{itemize}

We apply this retraining protocol to four detectors (BotBuster, RFS, LLM-based with \texttt{llama:70b}, and LLM-based with \texttt{mistral:24b}) across three benchmark datasets (Twibot-20, Cresci-2017, Caverlee-2011), yielding 12 detector-benchmark configurations. For each configuration, we measure relative F1-score improvement over the baseline detector that is initially pre-trained on the benchmark dataset and receives no incremental retraining. 

Figure~\ref{fig:rt_twibot} shows relative F1-score improvement across experimental days for each detector and retraining strategy, assuming Twibot-20 as benchmark pre-training dataset. See Appendix~\ref{app:incr_ret} for results with Cresci-2017 and Caverlee-2011 benchmarks. On Day 1, no improvement is observed for any strategy because both the baseline and retrained models are trained only on benchmark data (no experimental accounts are available yet for retraining).

\emph{Self-Supervision} yields marginal and inconsistent effects. The maximum improvement is 2.91\% (BotBuster, Day 2), while performance decreases up to -6.8\% (\texttt{llama:70b}, Day 3). This instability reflects the risk of reinforcing errors when retraining on the model's own predictions.

\emph{Ground-Truth Supervision} consistently improves performance, with gains reaching 17.4\% for BotBuster and RFS on Day 5. This confirms that access to correct labels, while an idealized scenario, provides reliable supervision for incremental retraining.

\emph{Human Supervision} generally yields improvements comparable to, and sometimes exceeding, \emph{Ground-Truth Supervision}. For \texttt{llama:70b} on Day 4, Human Supervision achieves 20\% improvement compared to 8.5\% for Ground-Truth. Similarly, for \texttt{mistral:24b} on Day 3, Human Supervision reaches 16.2\% versus 2.6\% for Ground-Truth. This counterintuitive result may be because \emph{Ground-Truth Supervision} only retrains on instances the model predicted correctly. When a detector has relatively few correct predictions, its retraining sample is small and less effective. \emph{Human Supervision}, by contrast, provides an independent labeling source that can include instances the model missed. However, it can also decrease performance, as observed for \texttt{llama:70b} on Days 2 and 3. This occurs when human reports are insufficient in quantity or unreliable in quality during early experimental days.

\emph{Overall, these findings support our hypothesis: human reports provide effective supervision for incremental retraining.} Human Supervision achieves improvements comparable to Ground-Truth Supervision without requiring access to ground-truth labels, offering a practical alternative for real-world deployment.

}

\section{Relevance to Social Platform Security}
Over 2000 years have passed since the publication of Sun Tzu's celebrated book, ``The Art of War''\cite{tzu1963art}. Yet, his statement that ``To know your Enemy, you must become your Enemy'' remains valid. It is widely known that social media companies use a mix of human ``content moderators'' and AI to detect bots and entities involved in CSIOs (and for many other purposes)\cite{lai2022human,levi2025ai}. To protect social platforms from malign covert social influence campaigns, they must ``become'' the enemy and understand how malign actors will leverage AI to run CSIOs in the future. As mentioned earlier, malign actors have already leveraged reinforcement learning to dynamically, adaptively, and autonomously evade 
static malware detectors
\cite{anderson2018learning,song2020mab}, create attack paths through a network
\cite{lee2025approach}, attack graph-neural network based detectors focused on node injection attacks\cite{sun2020adversarial}, and create fake reviews in online marketplaces\cite{zhang2023sockdef}. \cite{10.1145/3696410.3714729} provides a first message to social platforms that they need to anticipate and deter future CSIOs that leverage RL.

The findings of this paper, summarized in Table~\ref{tab:summary_f}, lead us to make several recommendations to social media companies.


\begin{reco}
Social platforms must \textbf{continuously monitor bot detection performance of both human reporters and AI bots} within their defensive arsenal.
\end{reco}
This recommendation is based on Hypothesis~\ref{hyp:f1-weighted} that shows that quality-based bot detection (in the human case) is better than merely counting the number of human reports that an account is a bot. 

\begin{reco}
Social platforms should \textbf{incorporate Hybrid (Human $+$ AI) bot detection within their bot detection infrastructure via either Meta Voting or Late Fusion}.
\end{reco}

This recommendation is based on Hypothesis~\ref{hyp:agg_str} (see also Table~\ref{tab:cv-aggregation}) which suggests that Meta-voting and Late Fusion achieve the best F1 score and Accuracy, respectively.  Such an incorporation could occur at several levels. For example, Hybrid bot detection could be implemented in order to determine which accounts should be reviewed by content moderators.  Alternatively, consider platforms with tiered moderation systems (e.g., ordinary moderators and senior trained supervisors). In the first round, Hybrid detection could combine user reports with AI predictions to flag accounts for junior moderators. In the second round, it could combine junior moderators' judgments with AI predictions to escalate difficult cases to senior supervisors.

\begin{reco}
\textbf{The human supervision strategy with quality-weighting should be used for retraining of AI-based bot detectors used by the social platform.}    
\end{reco}
This recommendation is based on Hypothesis~\ref{hyp:rt_s} which investigates three strategies for retraining: ground-truth supervision, self-supervision, and human-supervision and finds that human supervised training using quality-weighting avoids the need for access to ground truth labels and beats out both other strategies.

\begin{table}[t]
\centering
\caption{Summary of hypotheses and key findings.}
\label{tab:summary_f}

\begin{tabular}{cp{12.5cm}}
\toprule
\textbf{H\#} & \textbf{Key Finding}  \\
\midrule
\multicolumn{2}{l}{\textit{Demographic Characteristics}} \\
\cmidrule(lr){1-2}
1 & No significant gender difference in bot detection performance  \\
2 & Older participants ($\geq$35) outperformed younger ``digital natives''  \\
3 & Native English speakers showed 3$\times$ better detection than non-native speakers  \\
4 & Lower education (Bachelor's or below) associated with 8$\times$ better detection in terms of F1-score\\
5 & Social media platform usage (Twitter, Reddit) showed no systematic benefit  \\
\midrule
\multicolumn{2}{l}{\textit{Temporal \& Behavioral Factors}} \\
\cmidrule(lr){1-2}
6 & No learning effects observed over five days without feedback \\
7 & Active engagement with bots improves detection capabilities \\
8 & Exposure to bot behavior via received likes improves detection; follower relationships show no effect  \\
\midrule
\multicolumn{2}{l}{\textit{Collective Intelligence}} \\
\cmidrule(lr){1-2}
9 & Report frequency indicates bot likelihood, but 75\% of accounts were unreported  \\
10 & Quality-weighted aggregation outperforms count-based (+5.67\% F1, +9.46\% Acc.)  \\
\midrule
\multicolumn{2}{l}{\textit{Human vs. AI Detection}} \\
\cmidrule(lr){1-2}
11 & Humans achieve higher precision/accuracy; pre-trained AI achieves higher recall \\
12 & Hybrid ensembles outperform human and AI ensembles  \\
13 & Human supervision enables incremental retraining without ground-truth labels  \\
\bottomrule
\end{tabular}%

\end{table}

\section{Limitations}
Like any study, ours has several limitations. \textsf{(i)} Our U.S.-based participant sample may limit generalizability across different geographies. \textsf{(ii)} We evaluated AI detectors trained on benchmark datasets against experimental data without fine-tuning, creating realistic but challenging out-of-distribution conditions. While this mirrors many real-world deployments, new moderation systems may use incremental learning. \textsf{(iii)} The experimental DartPost platform differs from real-world social media, but must be used as ethical constraints and legal terms of use agreements prevent running CSIOs on real social platforms. \textsf{(iv)} The five-day experimental window may have been insufficient to capture learning effects emerging over longer periods, though running experiments over a longer duration present practical challenges. \textsf{(v)} Our analyses focus exclusively on RL-based adaptive bots \cite{10.1145/3696410.3714729}---the findings may not generalize to other bot types, campaigns, or coordination strategies. 

\section{Conclusion \& Future Work}

This study provides the first systematic investigation of human detection capabilities against RL-based adaptive bots operating in active CSIOs. Through controlled experimentation with 225 participants across five days, we tested 13 hypotheses spanning individual, social, and collective factors that shape bot detection performance. Our findings challenge some conventional assumptions: older participants outperformed younger digital natives, higher education correlated with worse detection, and social media platform usage showed no systematic benefit. Temporal analysis revealed no learning effects, while network centrality had no relationship with detection ability. Critically, active engagement with bots—not passive exposure—improved detection capabilities, suggesting that voluntary interaction enables development of behavioral recognition patterns. Collective intelligence analysis showed that report frequency is not a reliable indicator of bot likelihood. In contrast, accounting for reporter capability yields significantly better detection performance.


{ 

Benchmarking against three state-of-the-art AI bot detection algorithms revealed that the human ensemble performs better than any individual AI detector when both operate under realistic deployment conditions: humans received no feedback on their reports, while AI detectors were pretrained on benchmark datasets without fine-tuning on the experimental data. This comparison reflects scenarios where both humans and AI face novel, unseen influence campaigns. However, humans and AI detectors appear to complement each other, finding different bots.

We operationalized this complementarity by combining human reports with AI predictions. We investigated multiple aggregation strategies that integrate both human and AI detectors and show that the combination outperforms either in isolation: Meta Voting achieves the highest F1-score by capturing agreement patterns across all voter combinations, while Hybrid Late Fusion maximizes precision by incorporating high-performing human reports without inheriting their coverage limitations. Finally, human reports provide effective supervision for incremental retraining.

As mentioned previously, a strong defense requires the ability to ``know the enemy''\footnote{https://militairespectator.nl/artikelen/clausewitz-and-sun-tzu}. RL-based attacks are a likely wave in many settings including covert social influence operations. This paper analyzes human bot detection, AI-based bot detectors, and multiple Hybrid Combinations. We summarize our key actionable findings into recommendations that social platforms can potentially incorporate into their bot detection strategies, and which we hope will be useful to these social platforms in \emph{proactively} protecting against malign CSIO operators.
}

\bibliographystyle{ACM-Reference-Format}
\bibliography{sample-base}

\appendix

\section{Social Media Activity} \label{app:social_media_activity}

We examine the relationship between social media activity metrics and bot detection performance. Specifically, we analyze how Bot Engagement Ratio ($BER$) and Bot Exposure Ratio ($BXR$), measured separately for like and follow actions, correlate with users' detection capabilities in terms of precision, recall, and F1-score. $BER$ quantifies the proportion of a user's outgoing interactions directed toward bot accounts, calculated separately as $BER^{\text{like}}$ and $BER^{\text{follow}}$ These metrics reflect users' voluntary engagement with bot-generated content. 
$BXR$ quantifies the proportion of incoming interactions a user receives from bots, calculated separately as $BXR^{\text{like}}$ and $BXR^{\text{follow}}$. These metrics represent passive exposure to bot activity.

Table \ref{tab:app_social_activity} reports summary statistics of the regression models fitting users' precision, recall and F1 using $BER^{\text{like}}$, $BER^{\text{follow}}$, $BXR^{\text{like}}$, $BXR^{\text{follow}}$.

\begin{table*}[t]
\centering
\caption{Regression Analysis: Performance metrics as a function of $BER^{\text{like}}$, $BER^{\text{follow}}$, $BXR^{\text{like}}$, $BXR^{\text{follow}}$. }

\resizebox{.95\textwidth}{!}{
\begin{tabular}{r|ccc|ccc|ccc|ccc}
\toprule
& \multicolumn{3}{c|}{BER (like)} & \multicolumn{3}{c|}{BER (follow)} & \multicolumn{3}{c|}{BXR (like)} & \multicolumn{3}{c}{BXR (follow)} \\
\cmidrule(lr){2-4} \cmidrule(lr){5-7} \cmidrule(lr){8-10} \cmidrule(lr){11-13}
& Precision & Recall & F1 & Precision & Recall & F1 & Precision & Recall & F1 & Precision & Recall & F1 \\
\midrule
$R^2$     & $0.301$   & $0.225$ & $0.303$ & $0.221$ & $0.221$ & $0.223$ & $0.004$  & $0.0001$ & $0.002$   & $0.123$ & $0.111$  & $0.152$ \\
$\beta$   & $0.005$   & $0.007$ & $0.006$ & $0.004$ & $0.007$ & $0.005$ & $0.0007$ & $-0.001$ & $0.0006$  & $0.005$ & $0.0083$ & $0.0073$ \\
$p-$value & $0.00017^{***}$ & $0.001^{***}$ & $0.000071^{***}$ & $0.003^{**}$ & $0.003^{**}$ & $0.003^{**}$ & $0.701$  & $0.968$  & $0.788$   & $0.033^{*}$ & $0.044^{*}$  & $0.017^{*}$ \\
\bottomrule
\end{tabular}
}

\label{tab:app_social_activity}
\end{table*}

\section{Human vs AI Algorithms} \label{app:implementation}

We evaluated three representative bot detection methods spanning different technical paradigms and feature types:

\noindent\textbf{Hand-Crafted Feature Methods.} 
(i) \emph{RFS} \cite{9280525}  uses a random forest trained on behavioral and metadata features (follower counts, following counts, posting frequency) extractable from any social platform. This detector was employed in the original experiment and served as the adversarial target for RL-bot training.

(ii) \emph{BotBuster} \cite{ng2023botbuster} employs a multi-platform mixture of experts architecture combining three specialized components: post content analysis, user metadata features, and account-level posting patterns. We adapted the original implementation for our experimental setup, optimizing the number of experts, learning rates, and class weights to improve performance on our benchmark datasets.


\noindent\textbf{Generative Methods} 
(iii) \emph{LLM-based Detector} \cite{feng-etal-2024-bot} utilizes a prompt-based approach with in-context learning, where the model receives examples of bot and human accounts before classifying new instances. We evaluated four models from two families: \texttt{llama3.1:8b}, \texttt{llama3.1:70b}) and Mistral (\texttt{mistral:7b}, \texttt{mistral:24b}), using the official repository\footnote{\url{https://github.com/BunsenFeng/botsay}}. Table \ref{tab:llm_comparison} presents performance on the Twibot-20 dataset \cite{10.1145/3459637.3482019} across LLMs with varying numbers of in-context examples (0 vs. 4). With the exception of precision, \texttt{mistral:24b} consistently achieved the highest performance, and larger models (\texttt{llama3.1:70b}, \texttt{mistral:24b}) outperformed their smaller counterparts (\texttt{llama3.1:8b},  \texttt{mistral:7b}).

\begin{table}[t]
\centering
\caption{Performance comparison of LLM-based bot detectors on Twibot-20 with varying numbers of in-context examples: Precision, Recall and F1-score on the bot class, and overall Accuracy.  Bold indicates best performance, underline the first runner up.}

\begin{tabular}{r|c|cccc}
\toprule
\textbf{LLM} & \begin{tabular}[c]{@{}c@{}}\textbf{In-context}\\\textbf{Example}\end{tabular} & \textbf{Precision} & \textbf{Recall} & \textbf{F1} & \textbf{Accuracy} \\
\midrule
\multirow{2}{*}{\texttt{llama3.1:8b}} 
& 0 & 0.570 & 0.380 & 0.450 & 0.510 \\
& 4 & 0.500 & 0.330 & 0.400 & 0.460 \\
\midrule
\multirow{2}{*}{\texttt{llama3.1:70b}} 
& 0 & \textbf{0.825} & 0.228 & 0.357 & 0.556 \\
& 4 & 0.717 & 0.333 & 0.455 & 0.568 \\
\midrule
\multirow{2}{*}{\texttt{mistral-7b}} 
& 0 & 0.710 & 0.293 & 0.413 & 0.550 \\
& 4 & 0.560 & 0.301 & 0.561 & 0.492 \\
\midrule
\multirow{2}{*}{\texttt{mistral-24b}}
& 0 & \underline{0.732} & \underline{0.517} & \underline{0.606} & \underline{0.636} \\
& 4 & 0.729 & \textbf{0.615} & \textbf{0.668} & \textbf{0.669} \\
\bottomrule
\end{tabular}

\label{tab:llm_comparison}
\end{table}

\begin{table}
\centering
\caption{Performance on Benchmark Datasets. Bold indicates best performance, underline the first runner up.}
\label{tab:bot_benchmark}
\begin{tabular}{rcccccc}
\toprule
\textbf{Dataset} & \textbf{Model} & \textbf{Precision} & \textbf{Recall} & \textbf{F1} & \textbf{Accuracy} & \textbf{AUC} \\
\midrule
\multirow{4}{*}{Twibot-20} 
    & BotBuster & \textbf{0.756} & \textbf{0.986} & \textbf{0.856} & \textbf{0.819} & \textbf{0.857} \\
    & RFS       & 0.713 & \underline{0.915} & \underline{0.801} & \underline{0.746} & \underline{0.796} \\
    & LLM       & \underline{0.729} & 0.615 & 0.668 & 0.669 & $-$ \\

\midrule
\multirow{4}{*}{Cresci-2017}
    & BotBuster & \textbf{0.962} & \textbf{0.904} & \textbf{0.932} & \textbf{0.971} & \textbf{0.989} \\
    & RFS       & \underline{0.950} & \underline{0.851} & \underline{0.898} & \underline{0.957} & \underline{0.977} \\
    & LLM       & 0.582 & 0.464 & 0.516 & 0.807 & $-$ \\
\midrule
\multirow{4}{*}{Caverlee-2011}
    & BotBuster & \textbf{0.899} & \textbf{0.916} & \textbf{0.907} & \textbf{0.899} & \textbf{0.948} \\
    & RFS       & \underline{0.889} & {0.887} & \underline{0.888} & {0.860} & {0.929} \\
    & LLM       & 0.744 & 0.590 & 0.658 & 0.693 & $-$ \\
\bottomrule
\end{tabular}%

\end{table}

\subsection{Training and Evaluation}

All AI detectors except the LLM-based baseline were trained and validated using five-fold cross-validation on three benchmark datasets: Twibot-20~\cite{10.1145/3459637.3482019}, Cresci-2017~\cite{10.1145/3041021.3055135}, and Caverlee-2011~\cite{Lee_Eoff_Caverlee_2021}. The LLM-based detector relied on direct prompting with in-context examples rather than training, we tested four LLMs from two families: \texttt{llama3.1:8b}, \texttt{llama3.1:70b}) and Mistral (\texttt{mistral:7b}, \texttt{mistral:24b}). Table~\ref{tab:bot_benchmark} reports precision, recall, F1-score for the bot class, and overall accuracy and AUC across datasets. BotBuster achieved the best performance on all benchmark datasets.

For our comparative analysis with humans (Hypothesis \ref{hyp:h_better_m}), we applied these pre-trained models directly to the DartPost experimental data without additional training or fine-tuning. This evaluation setup mimics realistic deployment scenarios where detection algorithms encounter novel influence campaigns not represented in their training data. This setup also mirrors the human evaluation condition, where participants received no feedback during the experiment.

\begin{figure}
    \centering
    \includegraphics[width=.6\linewidth]{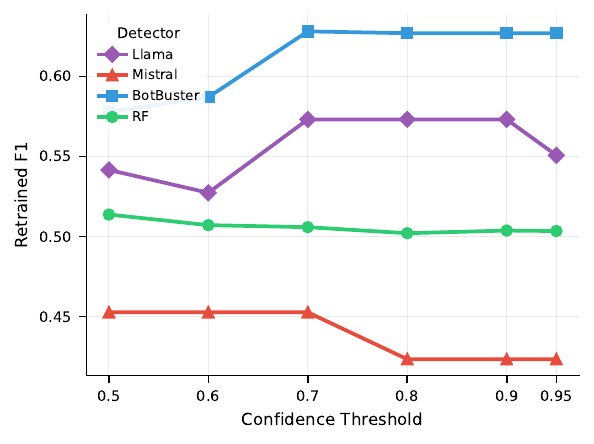}
    \caption{Effect of confidence threshold on Self-Supervision retraining. F1-score of retrained detectors (trained up to day 4, evaluated on day 5) on Twibot-20 as a function of the confidence threshold used to select retraining instances.}
    \label{fig:ss_threshold}
\end{figure}

{ 

\subsection{Incremental Retraining} \label{app:incr_ret}

Figures~\ref{fig:rt_cresci} and~\ref{fig:rt_caverlee} present relative F1-score improvement for detectors pretrained on Cresci-2017 and Caverlee-2011, respectively. Results are consistent with those observed for Twibot-20 (Figure~\ref{fig:rt_twibot}): \emph{Ground-Truth Supervision} consistently improves performance except for \texttt{llama:70b}, while \emph{Human Supervision} achieves comparable gains but may still decrease performance in some cases.

We also observe an effect of benchmark dataset on retraining effectiveness. Pretraining on Cresci-2017 (Figure~\ref{fig:rt_cresci}) yields lower performance improvements for \texttt{llama:70b}, \texttt{mistral:24b}, and BotBuster compared to other benchmarks. This likely reflects a language mismatch: Cresci-2017 is multilingual while our experiment was conducted in English. Supporting this interpretation, RFS, which does not rely on content features, still achieves substantial improvements on this benchmark (up to 31.1\% on Day 5 for Ground-Truth Supervision and 21.9\% on Day 5 for Human Supervision).

\begin{figure*}
    \centering
    \includegraphics[width=\textwidth]{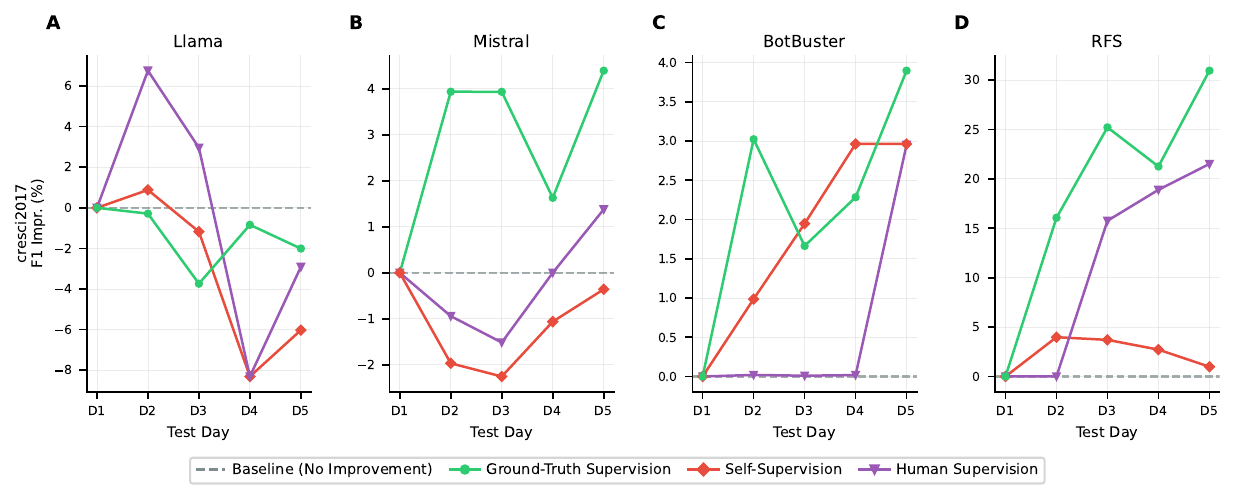}
    \caption{Relative F1-score improvement over baseline (no retraining) across experimental days for four detectors pretrained on Cresci-2017. Each subplot shows one detector. \emph{Ground-Truth Supervision} retrains on correct predictions, \emph{Self-Supervision} retrains on high-confidence predictions (threshold 0.7), and \emph{Human Supervision} retrains on human reports.}
    \label{fig:rt_cresci}
\end{figure*}

\begin{figure*}
    \centering
    \includegraphics[width=\textwidth]{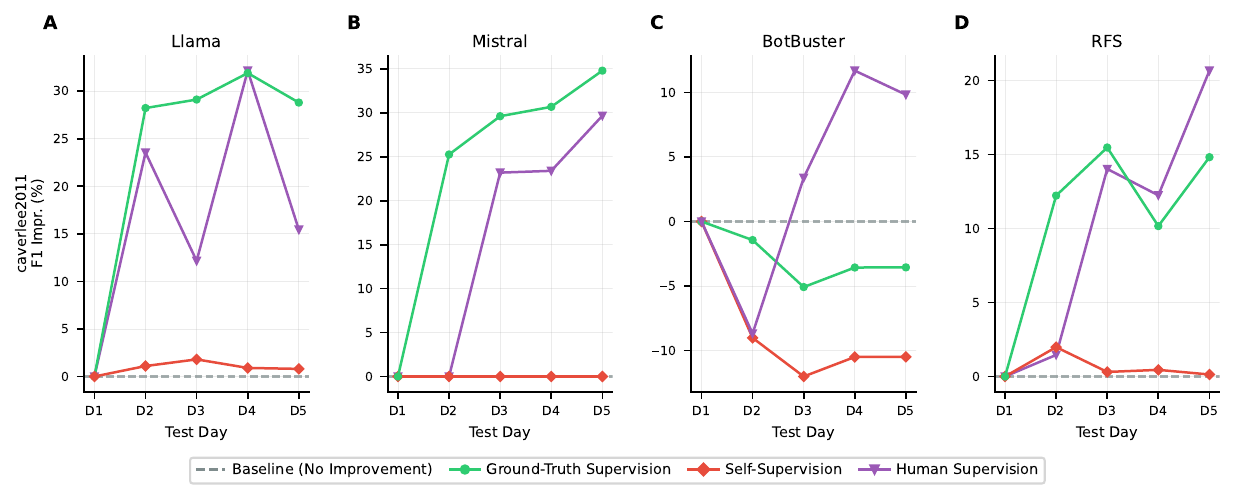}
    \caption{Relative F1-score improvement over baseline (no retraining) across experimental days for four detectors pretrained on Caverlee-2011. Each subplot shows one detector. \emph{Ground-Truth Supervision} retrains on correct predictions, \emph{Self-Supervision} retrains on high-confidence predictions (threshold 0.7), and \emph{Human Supervision} retrains on human reports.}
    \label{fig:rt_caverlee}
\end{figure*}

\subsubsection{Self-Supervision Threshold Selection} \label{app:ss_ret}

In the \emph{Self-Supervision} retraining strategy, detectors are retrained on instances they predicted as bots with confidence exceeding a threshold. This approach requires no access to ground-truth labels, making it attractive for fully automated deployment. However, the choice of threshold involves a trade-off: higher thresholds isolate predictions where the model is most confident, but this is not necessarily beneficial as retraining on these instances may reinforce systematic errors. Additionally, higher thresholds reduce the number of instances available for retraining, limiting the model's exposure to new behavioral patterns.
We empirically evaluated thresholds from 0.5 to 0.95 across all four detectors. Figure~\ref{fig:ss_threshold} presents F1-scores for models trained up to day 4 and evaluated on day 5 using the Twibot-20 benchmark.

Results show that a threshold of 0.7 yields the best or equivalent F1-score across all detectors. For RFS, performance remains stable regardless of threshold, suggesting this detector's confidence scores are not informative for instance selection. For Mistral, performance degrades at thresholds above 0.7, indicating that high-confidence predictions may reinforce errors. For Llama and BotBuster, performance plateaus at 0.7 and remains stable at higher thresholds. Given these results, we select 0.7 as it achieves optimal performance while maximizing the number of instances available for retraining.

}

\end{document}